\documentclass[10pt,twocolumn]{article}

\usepackage[
  a4paper,
  top=0.76in,        
  bottom=1in,
  left=1in,
  right=1in,
  headheight=14pt,   
  includehead,
  includefoot
]{geometry}

\usepackage{placeins}
\usepackage{xurl} 
\usepackage{float}
\usepackage{graphicx}   
\usepackage{amsmath, amssymb}
\usepackage{cite}
\usepackage{xcolor}
 \usepackage{tabularx}
\usepackage{url}

\usepackage{authblk}

\usepackage[colorlinks=true,
            linkcolor=black,
            citecolor=black,
            urlcolor=black,
            bookmarks=true]{hyperref}

\usepackage{titlesec}
\usepackage{titling}
\usepackage{setspace}
\usepackage{lmodern}
\usepackage{placeins}
\titleformat{\section}
  {\normalfont\large\bfseries}
  {\thesection}
  {1em}{}
\titlespacing*{\section}
  {0pt}{1.0ex plus 0.5ex minus 0.5ex}{0.8ex plus 0.2ex}
\titleformat{\subsection}
  {\normalfont\normalsize\bfseries}
  {\thesubsection}
  {1em}{}
\titlespacing*{\subsection}{0pt}{1pt}{3pt}

\titleformat{\subsubsection}
  {\normalfont\normalsize\bfseries}
  {\thesubsubsection}
  {1em}{}

\title{\vspace{-5em}\textbf{Lung Cancer Survival Prediction Using Machine Learning and Statistical Methods}}

\author{
    \begin{minipage}[t]{0.48\textwidth}
        \centering
        \textbf{Varun Vishwanathan Nair}\\
        \small{Department of Mathematical Sciences}\\
        \small{Auckland University of Technology}\\
        \small{Auckland, New Zealand}\\
        \footnotesize\href{mailto:yfg2694@autuni.ac.nz}{yfg2694@autuni.ac.nz}
    \end{minipage}
    \hfill
    \begin{minipage}[t]{0.48\textwidth}
        \centering
        \textbf{Victor Miranda Soberanis}\\
        \small{Department of Mathematical Sciences}\\
        \small{Auckland University of Technology}\\
        \small{Auckland, New Zealand}\\
        \footnotesize\href{mailto:victor.miranda@aut.ac.nz}{victor.miranda@aut.ac.nz}
    \end{minipage}
}

\begin{document}
\maketitle

\centerline{\textbf{Abstract}}

Lung cancer remains one of the leading causes of cancer-related mortality, yet most survival models rely only on baseline factors and overlook post-treatment variables that reflect disease progression. To address this gap, we applied Cox Proportional Hazards and Random Survival Forests, integrating baseline features with post-treatment predictors such as progression-free interval (PFI.time) and residual tumor status. The Cox model achieved a concordance index (C-index) of 0.90, while the RSF model reached 0.86, both outperforming previous studies. Beyond statistical gains, the integration of post-treatment variables provides oncologists with more clinically meaningful and reliable survival estimates. This enables improved treatment planning, more personalized patient counseling, and better-informed follow-up strategies. From a practical standpoint, these results demonstrate how routinely collected clinical variables can be transformed into actionable survival predictions.\\
\textbf{Keywords-}
Lung Cancer Survival, Cox Proportional Hazards Model, Random Survival Forests, Dynamic Survival Modeling, Kaplan-Meier Analysis. 

\section{Introduction}
Lung cancer remains one of the leading causes of cancer-related mortality worldwide, and predicting patient survival continues to be a major challenge. Traditional prognostic models rely primarily on static baseline variables such as age, gender, and tumor stage. While useful, these models often fail to capture the dynamic clinical course of the disease, where treatment response, residual tumor, or progression-free survival can drastically alter outcomes. As a result, existing approaches provide only limited accuracy and personalization in survival estimation, reducing their value for real-world clinical decision-making\cite{C.M.Caruso2024}.

Statistical methods such as the Kaplan–Meier estimator and Cox Proportional Hazards (CPH) model have long formed the foundation of survival analysis. However, Kaplan–Meier is descriptive and does not account for covariates, while CPH assumes proportional hazards that may not hold in heterogeneous cancer populations. More recent machine learning methods, including Random Survival Forests (RSF), can model complex, non-linear relationships and manage censored data\cite{M.K.Goel2010}. Yet, even these are often applied only to baseline features, overlooking post-treatment variables that provide critical insight into evolving patient risk\cite{I.Etikan2017}.

This study addresses that gap by integrating both baseline and post-treatment predictors into survival modeling. In particular, we incorporate progression-free interval (PFI.time) and residual tumor status alongside age and gender to capture changes in disease trajectory after therapy. We employ two complementary approaches: CPH for interpretability through hazard ratios, and RSF for flexible modeling of non-linear interactions. Using the TCGA lung adenocarcinoma (LUAD) dataset, we evaluate the predictive performance of these models and compare them against established benchmarks in the field.

Following previous evidence, our results strongly suggest that incorporating post treatment predictors into survival models leads to a substantial improvement in performance. Compared with baseline-only approaches reported in earlier studies, both Cox Proportional Hazards and Random Survival Forests achieved higher concordance indices, with Cox reaching 0.90 and RSF 0.86. More importantly, the inclusion of progression-free interval and residual tumor status produces survival estimates that are not only statistically robust but also clinically relevant, aligning more closely with how oncologists assess patient outcomes. These contributions illustrate that routinely collected clinical data can be repurposed into actionable survival predictions, offering a meaningful step forward over conventional prognostic approaches.

\section{Literature Review}

\subsection{Conventional Statistical and Prognostic Models}
Survival analysis in oncology has long relied on statistical models such as the Kaplan–Meier (KM) estimator and the Cox Proportional Hazards (CPH) model. The KM estimator is a non-parametric method widely used to estimate survival probabilities and plot survival curves, enabling comparison across patient groups such as treatment types, stages, or biomarkers. Its strength lies in handling censored data, making it a standard tool for exploratory analysis in clinical studies \cite{I.Etikan2017}, \cite{M.K.Goel2010}. However, KM is purely descriptive and cannot incorporate covariates, limiting its use for predictive modeling or multivariate settings in personalized medicine.

The Cox model was developed as a semi-parametric alternative, allowing covariates such as age, gender, tumor size, and biomarkers to be included when estimating hazard ratios the relative risk of an event at a given time. This balance of flexibility and interpretability has made it a cornerstone in prognostic modeling, with studies \cite{J.Zhu2022}, \cite{S.Khatua2024} showing its effectiveness in predicting lung cancer outcomes and building clinical nomograms using variables like ECOG status, tumor diameter, and serum biomarkers.

Despite its advantages, the CPH model assumes proportional hazards, meaning covariate effects remain constant over time. In reality, especially for heterogeneous populations like lung cancer, treatment effects and disease progression often vary, violating this assumption. Moreover, Cox has limited ability to capture non-linear relationships or complex interactions, reducing its utility in high-dimensional or dynamically evolving clinical settings.

Baseline prognostic models have also been widely applied in oncology. For example, \cite{E.Chow2008} developed and validated a simplified three-factor model incorporating cancer site, metastasis site, and performance status, which showed strong internal and external validity. While clinically practical, such baseline-only models cannot reflect disease progression over time, underscoring the need for approaches that integrate post-treatment predictors.

\subsection{Machine Learning Methods: Random Survival Forests and Classification Models}

Recent advances in machine learning (ML) have produced flexible survival models capable of capturing non-linear effects, modeling interactions, and handling high-dimensional data. Among these, Random Survival Forests (RSF) have gained prominence \cite{A.Groji2025}. RSF is an ensemble method that builds multiple survival trees on bootstrapped samples and aggregates predictions, enabling robust results without proportional hazards or parametric assumptions. It naturally manages missing values and complex covariate interactions. Studies have shown RSF to outperform Cox Proportional Hazards (CPH), particularly in heterogeneous datasets \cite{J.A.Bartholomai2018}, and in a lung cancer cohort, RSF achieved a higher concordance index (C-index) than Cox, indicating better predictive alignment \cite{S.Khatua2024}.

Classification-based models have also been applied by grouping patients into categories (e.g., $<$6 months, 6–24 months, $>$24 months). These models simplify survival analysis and can support triage and stratification decisions \cite{F.Y.Dao2020}. When combined with regression or calibration, they help segment populations and guide treatment. However, discretization of survival times causes information loss, performance depends heavily on arbitrary cutoffs, and generalizability across settings is limited.

Despite improved predictive accuracy, many ML methods are criticized as black-box approaches \cite{Z.Shi2025}, \cite{L.Ter-Minassian2024}. Without post hoc interpretability tools such as SHAP or LIME, their clinical adoption remains challenging because predictions cannot be easily explained in evidence-based medicine.

\subsection{Deep Learning Models}

Deep learning (DL) has recently gained traction in survival analysis due to its ability to model complex relationships and manage incomplete clinical data. In the study \cite{C.M.Caruso2024} a novel transformer-based architecture was proposed to predict overall survival in non-small cell lung cancer (NSCLC) patients. Unlike traditional approaches, this model directly handles missing values without requiring imputation by encoding temporal and semantic relationships within clinical features. The architecture integrates both static and time-dependent variables and learns latent representations that enhance survival prediction, even in the presence of incomplete or irregularly sampled patient data. This methodology reflects a shift toward more flexible and robust modeling frameworks in clinical prognostics.

Another promising approach is DeepHit, introduced in \cite{C.Astley2023}\cite{Y.H.Lai2020}, which uses a deep neural architecture to estimate survival probabilities directly. This model accounts for competing risks and generates a full survival distribution for each patient, improving flexibility and accuracy. Tools like LIME (Local Interpretable Model-agnostic Explanations) have also been applied to these architectures, providing local explanations for individual predictions and partially addressing the interpretability concern.

Despite these advancements, DL models face significant practical limitations. They require large volumes of high-quality, annotated training data, which are often unavailable in clinical practice due to privacy issues, inconsistent data standards, and sparse follow-up records. Furthermore, training deep models involves substantial computational resources and expertise, making their integration into clinical workflows difficult. Even with emerging explainability techniques, most deep learning models still struggle to achieve the transparency needed for clinical accountability. As a result, adoption remains limited to research settings or highly resourced institutions.
\subsection{Dynamic Modeling Approaches}
In recent years, dynamic survival modeling has emerged as a critical area of innovation. Unlike static models, which rely solely on baseline covariates, dynamic models incorporate time-updated information, reflecting the patient’s evolving clinical trajectory. These models are particularly well-suited for diseases like lung cancer, where treatment response, biomarker levels, and functional status can change significantly over time. One of the foundational contributions to this area was by \cite{R.Henderson2000}, who demonstrated that joint modeling of longitudinal measurements and survival outcomes provides more realistic and robust estimates of patient risk compared to baseline-only models.

Dynamic modeling techniques include methods that update survival probabilities as new information becomes available for instance, tumor recurrence, treatment changes, or follow-up imaging results. Studies like \cite{S.Khatua2024} and \cite{C.M.Caruso2024} have developed such models using time-series clinical data and recurrent neural networks, allowing real-time survival prediction and continuous risk stratification. These models align closely with the real-world nature of patient care, where clinicians adjust decisions based on a patient’s latest condition rather than static baseline features.

However, deploying dynamic models in practice poses multiple operational and technical challenges. The most significant is the need for longitudinal data collection, which is not standardized across healthcare systems. Many electronic health records lack the consistency or structure required for effective time-series modeling. Moreover, real-time prediction systems must be continuously retrained and recalibrated as new data becomes available, demanding robust infrastructure and cross-functional collaboration between data scientists and clinicians. The cost and complexity of implementing such systems often outweigh perceived benefits, especially in resource-limited settings.

\subsection{Model Evaluation and Clinical Utility}
Evaluating survival models requires balancing statistical performance with clinical usefulness. Metrics such as the concordance index (C-index) remain standard for assessing discrimination in censored survival data, as they quantify how well a model distinguishes between patients with different outcomes. However, as \cite{M.Assel2017} argue, commonly used measures like the C-index and Brier score focus on statistical accuracy but do not capture whether predictions improve decision-making in clinical practice. This limitation is particularly relevant in oncology, where the ultimate value of a survival model lies in its ability to inform treatment planning, guide follow-up strategies, and support personalized patient counseling. Accordingly, while this study reports conventional metrics such as the C-index, it also emphasizes the practical contribution of integrating post-treatment variables, demonstrating how statistical improvements translate into more clinically meaningful survival predictions.
\section{Methodology}

\subsection{Kaplan Meier Estimator}

The Kaplan-Meier estimator is a non-parametric statistic used to estimate the survival function from observed survival times. It is particularly useful for visualizing survival probabilities and comparing different groups (e.g., based on gender or age). In this study, KM curves were generated for the overall dataset, and stratified analyses were conducted by gender and age group to explore subgroup survival patterns.

The survival probability at time t is denoted by S(t), is given by the product-limit formula

\begin{center}
\begin{tabular}{@{}l@{\hspace{2cm}}r@{}}
$S(t) = \displaystyle\prod_{t_i \leq t} \left( 1 - \frac{d_i}{n_i} \right),$ &  (1)
\end{tabular}
\end{center}

where:\\
- \( t_i \) is the time of the \( i \)-th event (e.g., death)\\
- \( d_i \) is the number of events at \( t_i \)\\
- \( n_i \) is the number of individuals at risk just prior to \( t_i \).\\

This estimator accounts for right-censored data, where the exact time of the event is unknown for some individuals due to loss to follow-up or study end\cite{I.Etikan2017}\cite{M.K.Goel2010}.\\

\subsection {Cox Proportional Hazard Modelling:}

\normalsize
The Cox Proportional Hazards model is a semi-parametric method that models the hazard function, allowing for the inclusion of multiple covariates.\cite{S.Salerno2023} It assumes that the effect of each covariate is multiplicative with respect to the baseline hazard and remains constant over time (i.e., proportional hazards assumption).

The hazard function for subject i at time t is :\\
\setcounter{equation}{1} 

\begin{center}
\begin{equation}
h_i(t) = h_0(t)\exp(\beta_1 x_{i1} + \beta_2 x_{i2} + \cdots + \beta_p x_{ip})
\tag{2}
\end{equation}
\end{center}

where:
- \( h_i(t) \) is the hazard for individual \( i \) at time \( t \)\\
- \( h_0(t) \) is the baseline hazard function,\\
- \( x_{ij} \) are the covariates for individual \( i \),\\
- \( \beta_j \) are the coefficients corresponding to each covariate.\\

This model provides interpretable hazard ratios for each predictor and is widely used in clinical research due to its simplicity and effectiveness in analyzing censored data\cite{L.Pu2024}.\\

\subsection{Random Survival Forest:}

Random Survival Forests (RSF) are an ensemble learning method that extends decision trees to time-to-event data. RSFs can handle high-dimensional datasets, non-linear relationships, and complex interactions without making assumptions about the underlying hazard distribution.
\[
\hat{S}(t \mid X_i)= P(T > t \mid X = X_i), \hspace{2em} ~(3)
\]

where:

T is the time-to-event (e.g., death, failure)

X 
i
  is the feature vector for individual 

\(\hat{S}(t \mid X_i)\) is the estimated survival probability at time \(t\).\\

In this study, the RSF model was trained using the scikit-survival package, with survival labels encoded using Surv.from dataframe(). The model was optimized using parameters such as the number of trees, minimum samples per split, and maximum features. RSF provided risk predictions and a concordance index (C-index) was used to evaluate performance.

By integrating classical statistical techniques with machine learning-based RSF, this study allows for both interpretability and improved predictive power. The three models were subsequently evaluated and compared based on performance metrics and clinical interpretability.

\subsection{C-index}
\begin{equation}
C = \frac{\sum_{i < j} \mathbf{1}\big( \hat{h}_i < \hat{h}_j \,\wedge\, T_i < T_j \big)}
{\sum_{i < j} \mathbf{1}(T_i < T_j)}
\tag{4}
\end{equation}

where:

-$C$ is the concordance index

-$\hat{h}_i$ and $\hat{h}_j$ are predicted risk scores for individuals

$i$ and $j$, 
$T_i$ and $T_j$ are their observed survival times, 
and $\mathbf{1}(\cdot)$ is the indicator function.

To assess the relative performance of the survival models, this study employs a combination of evaluation metrics. The Concordance Index (C-index) serves as the primary metric, reflecting each model’s ability to correctly rank patient survival times.

\subsection{ROC Curve and AUC}

The Receiver Operating Characteristic (ROC) curve is defined in terms of the
True Positive Rate (TPR) and False Positive Rate (FPR):

\begin{equation}
\text{TPR} = \frac{TP}{TP + FN}, \qquad
\text{FPR} = \frac{FP}{FP + TN}
\tag{5a}
\end{equation}

where:
$TP$ = true positives, $FN$ = false negatives, 

$FP$ = false positives, and $TN$ = true negatives.

The ROC curve is obtained by plotting $\text{TPR}$ against $\text{FPR}$ as the classification threshold varies.

The area under the ROC curve (AUC) is then given by:

\begin{equation}
\text{AUC} = \int_0^1 \text{TPR}\big(\text{FPR}^{-1}(x)\big) \, dx
\tag{5b}
\end{equation}

where AUC represents the probability that the model ranks a randomly chosen
positive instance higher than a randomly chosen negative instance.

Additionally, ROC (Receiver Operating Characteristic) curves are used to assess the discriminative ability of the models at specific time thresholds (1000 days), providing a visual and quantitative basis for comparison. These metrics are complemented by an analysis of model interpretability and underlying assumptions. 

While the primary focus of this study is on survival modeling using time-to-event data, preliminary experiments with classification-based approaches on a separate cancer dataset were conducted during the early phases of model exploration. These models, however, yielded poor results achieving around 50 percent accuracy with weak precision and recall highlighting their unsuitability for predicting censored survival outcomes. This outcome further reinforced the decision to adopt dedicated survival analysis techniques such as Cox Proportional Hazards and Random Survival Forests for this study.
\section{Implementation}
\subsection{Data collection:}
The clinical data for this study were obtained from the publicly available TCGA LUAD (Lung Adenocarcinoma) cohort via the UCSC Xena Browser. Two key datasets were downloaded and the coding part of the project was done using Google colab:

LUNG survival.txt – containing survival-related variables, including overall survival (OS), OS time, progression-free interval (PFI), and vital status.
\href{https://xenabrowser.net/datapages/?dataset=survival%2FLUNG_survival.txt&host=https%3A%2F%2Ftcga.xenahubs.net&removeHub=http%3A%2F%2F127.0.0.1%3A7222}{LUNG\_survival.txt dataset}

LUAD clinicalMatrix.json – containing a range of clinical features such as age at diagnosis, gender, tumor stage, treatment response, and additional pathological and molecular indicators.
\href{https://xenabrowser.net/datapages/?dataset=TCGA.LUAD.sampleMap%2FLUAD_clinicalMatrix&host=https%3A%2F%2Ftcga.xenahubs.net&removeHub=https%3A%2F%2Fxena.treehouse.gi.ucsc.edu%3A443}{LUAD\_clinicalMatrix.json dataset} .

\subsection{Data Preprocessing:}

Preprocessing was a critical step in preparing the dataset for survival analysis and ensuring methodological rigor. The dataset was created by merging two publicly available sources from the TCGA LUAD cohort via the UCSC Xena Browser using sample ID as the common key. This integration yielded approximately 1,300 patient records and 146 variables, including demographic, clinical, and survival data. In line with ethical standards, all data were de-identified and publicly accessible, eliminating concerns around privacy or consent. The preprocessing process began with the removal of completely empty columns, followed by inspection of missing data. Rows missing values in critical target variables (OS.time and OS) were dropped to preserve label integrity, while missing predictor values were imputed using median imputation via scikit-learn's SimpleImputer to retain usable data without imposing unrealistic distributional assumptions.

To prepare the dataset for machine learning algorithms, categorical variables were encoded appropriately: gender was label-encoded, and the residual tumor variable was one-hot encoded to treat each category independently. A correlation heatmap was generated to support feature selection by identifying variables with stronger linear relationships to OS.time and minimizing multicollinearity. Predictor selection was guided by clinical relevance, data completeness, and exploratory data analysis. Features such as PFI.time, age at diagnosis, days to new tumor event, gender, and residual tumor were retained due to their relevance and acceptable data quality, while others with weak correlation, excessive missingness, or limited clinical justification were excluded to improve model clarity and generalizability. The final structured dataset was encoded, imputed, and filtered, then split into training and test sets using 80-20 stratification. This comprehensive preprocessing pipeline ensured consistency, minimized bias from data loss, and upheld the ethical and technical standards required for clinical survival modeling.

After identifying extreme survival durations through the initial preprocessing steps, outliers in the OS.time variable were removed using the Interquartile Range (IQR) method. Specifically, any values falling below Q1  1.5 × IQR or above Q3 + 1.5 × IQR were excluded.  Moreover, removing outliers helped stabilize estimates of central tendency (mean, median) and variance, thereby improving the reliability of modeling and interpretation. While advanced models like Random Survival Forests can tolerate outliers, this step was important for ensuring consistent preprocessing across all methods and highlighting how much of the original distribution was driven by a small number of extreme cases.

The dataset exhibits a noticeable imbalance in the binary event indicator (OS), where a value of 1 denotes patients who experienced the event (death) and 0 indicates censored cases (i.e., patients lost to follow-up or still alive at the time of analysis). Approximately 64\% of patients in the dataset are censored, while 36\% have experienced the event. This level of censoring is common in clinical survival datasets and highlights the importance of choosing models and evaluation strategies that appropriately account for censored observations.

This imbalance is common in real-world clinical datasets, particularly in longitudinal studies, where many patients may not have reached the event endpoint (death) by the time of data extraction. Contributing factors include limited follow up time for recently enrolled patients, improved treatment outcomes, and participant dropout. Although such levels of censoring are typical, they can negatively impact model performance if not properly addressed. To mitigate this, the study employed censoring-aware survival models such as Kaplan-Meier, Cox Proportional Hazards, and Random Survival Forests, which are explicitly designed to handle incomplete outcome data without introducing bias. In addition to handling censoring, a correlation analysis was conducted to examine the linear relationships among numeric variables and assess the degree of multicollinearity  a critical factor for model stability and interpretability, especially in regression-based models. The analysis revealed that most numeric predictors were only weakly to moderately correlated, indicating that multicollinearity was not a major concern. This insight guided feature selection, helping retain variables that were both clinically relevant and statistically independent. Specifically, predictors such as PFI.time (progression-free interval), age at initial diagnosis, and days to new tumor event after treatment were selected based on their significance and low inter correlation with other variables, ensuring that the final model inputs remained robust and informative.

\section{Results}
\subsection{Kaplan Meier Survival }

\begin{figure}[!t]
   \centering
    \includegraphics[width=0.40\textwidth]{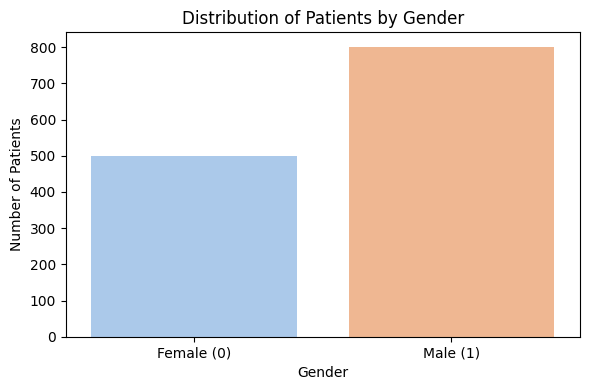}
   
\caption{\small Gender wise distrbution of population }
    \label{fig:Gender_Distribution_Diagram}
\end{figure}

\begin{figure}[!t]
   \centering 
    \includegraphics[width=0.45\textwidth]{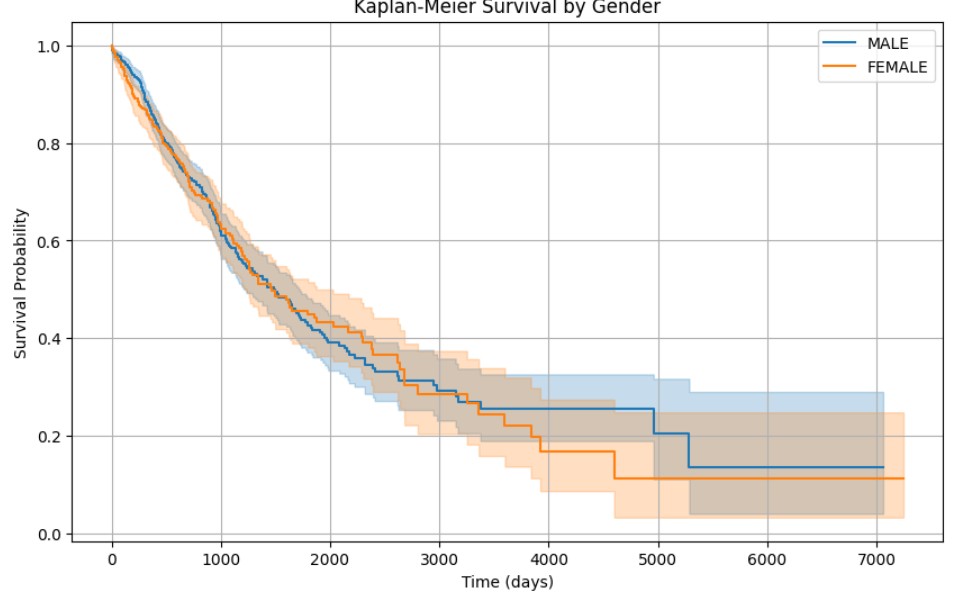}
   
\caption{\small Kaplan-Meier Survival graph by gender }
    \label{fig:KaplanMeier-diagram}
\end{figure}
\vspace{-1em}

The Kaplan-Meier survival curves stratified by gender (Fig 2), with males represented in blue and females in orange. The plot illustrates the estimated survival probabilities over time, starting from 100\% and decreasing as death events occur.While both curves follow a similar declining pattern, male patients appear to exhibit slightly better long-term survival, particularly beyond 4000 days. This may be influenced, at least in part, by the unequal gender distribution observed in the dataset. As shown in the corresponding gender bar plot (Figure 1), there are more male patients (801) than female patients (498). This imbalance in group sizes can affect the survival estimates, especially in the tail of the distribution where fewer female patients remain under observation.

The shaded areas around each curve represent 95\% confidence intervals, which grow wider as fewer patients remain at risk. Although minor differences in survival are observed between genders, the overall shape of the curves suggests broadly similar survival patterns, with the gender imbalance potentially contributing to the slight divergence seen in later stages.This indicates that while gender may play a modest role, other factors likely have a stronger influence on long-term survival outcomes\cite{M.Othus2012}.

The Kaplan–Meier curves provided an initial descriptive view of survival outcomes. Stratification by gender suggested broadly similar survival probabilities, with minor long-term differences potentially influenced by unequal sample sizes between males and females. Age-based stratification showed that older patients (61–100 years) appeared to have slightly higher long-term survival compared to younger patients (0–60 years), though this trend may reflect cohort imbalance rather than a true biological effect. These results illustrate how KM curves can highlight subgroup patterns, but also reinforce the need for multivariable approaches such as Cox regression and RSF to obtain more robust and clinically meaningful predictors of survival.

\begin{figure}[!t]
   \centering
    \includegraphics[width=0.45\textwidth]{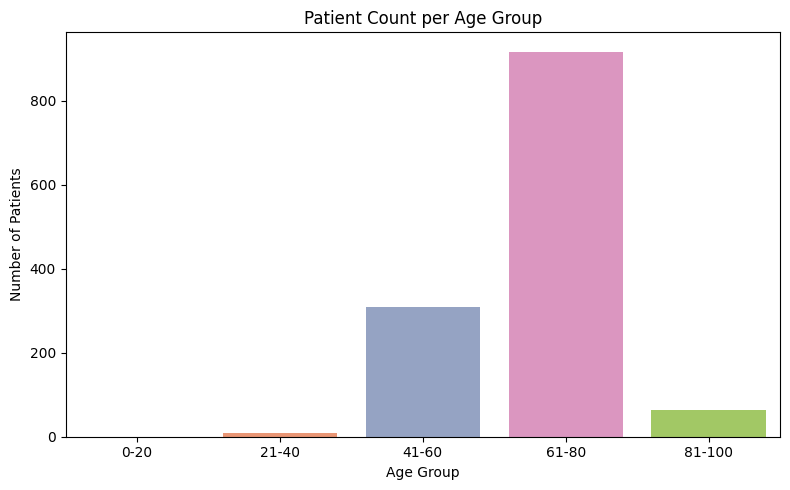}
   
\caption{\small Distribution of Age Group }
    \label{fig:Barplot-diagram}
\end{figure}
\vspace{-0.1em}

\begin{figure}[!t]
   \centering
    \includegraphics[width=0.45\textwidth]{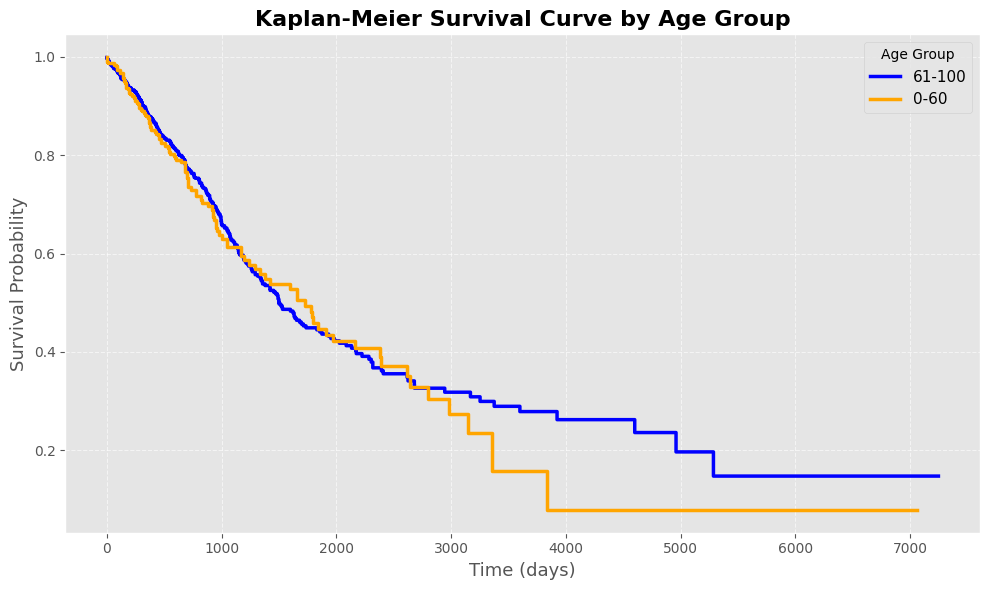}  
\caption{ \small Kaplan Meier Survival Graph by Age Group }
    \label{fig:KaplanMeier-diagram}
\end{figure}
\vspace{-0.3em}
Figure 3 shows the distribution of patients across predefined age groups. The majority of patients fall within the 61–80 age group, accounting for the highest number of cases (917 patients). This is followed by the 41–60 group, which includes 310 patients. The 81–100 group contributes 64 patients, while only 8 patients are in the 21–40 group, and none are in the 0–20 group.

This distribution reflects a skew toward older age groups,  as the incidence of disease increases with age. The low representation in the younger age brackets suggests that age-stratified analyses should focus on splitting age group in 2 categories 0-60 and 61-100 for a better stratified graph, where the data is more concentrated and reliable for survival modeling.

The (Figure 4) displays Kaplan-Meier survival curves for two age groups: 0–60 years (yellow) and 61–100 years (blue). These curves estimate the probability of survival over time (in days) for each group, starting at 100\% and declining as events (deaths) occur.
\begin{table*}[!t]
\centering
\caption{Cox Proportional Hazards Model Coefficients}
\label{tab:cox_results}
\begin{tabular}{lccccccc}
\hline
\textbf{Variable} & \textbf{coef} & \textbf{exp(coef)} & \textbf{se(coef)} & \textbf{95\% CI (coef)} & \textbf{95\% CI (HR)} & \textbf{z} & \textbf{p} \\
\hline
PFI.time & -0.00 & 1.00 & 0.00 & -0.00 -- -0.00 & 1.00 -- 1.00 & -18.33 & $<$0.005 \\
days\_to\_new\_tumor\_event & 0.00 & 1.00 & 0.00 & -0.00 -- 0.00 & 1.00 -- 1.00 & 0.74 & 0.46 \\
age\_at\_diagnosis & 0.01 & 1.01 & 0.01 & -0.00 -- 0.02 & 1.00 -- 1.02 & 1.26 & 0.21 \\
gender\_encoded & -0.23 & 0.80 & 0.08 & -0.39 -- -0.07 & 0.68 -- 0.94 & -2.76 & 0.01 \\
residual\_tumor\_R1 & -0.29 & 0.75 & 0.29 & -0.87 -- 0.29 & 0.42 -- 1.33 & -0.99 & 0.32 \\
residual\_tumor\_R2 & 0.46 & 1.58 & 0.45 & -0.43 -- 1.34 & 0.65 -- 3.82 & 1.01 & 0.31 \\
residual\_tumor\_RX & 0.32 & 1.38 & 0.26 & -0.18 -- 0.82 & 0.83 -- 2.28 & 1.25 & 0.21 \\
\hline
\end{tabular}
\end{table*}
In the initial stages, both groups show similar survival probabilities. However, beyond approximately 3000 days, a notable divergence appears: the older age group (61–100) maintains a higher survival probability than the younger group (0–60). By the end of the follow-up period, the survival curve for the younger group levels off at a lower point, indicating poorer long-term survival compared to older patients.

Although unusual, this trend may be influenced by factors such as disease severity, treatment differences, or stage at diagnosis that differ between age groups. Additionally, the larger number of patients in the 61–100 group may have contributed to the stability of their survival estimates. These results suggest that older patients in this cohort had better long-term survival, warranting further investigation into contributing clinical factors.

\subsection{Cox Proportional Hazard Model}

The Cox proportional hazards model revealed that progression-free interval (PFI.time) was the strongest predictor of survival, with longer PFI associated with significantly reduced hazard of death ($p < 0.005$). Although the hazard ratio rounded to 1.00 due to the scale of measurement in days, the strong statistical significance confirms that even small increments in PFI time translate into meaningful survival benefits, consistent with its role as a critical post-treatment variable. Gender also emerged as a significant factor, with males showing a 20\% lower hazard compared to females (HR = 0.80, p = 0.01). Residual tumor status demonstrated clinically consistent directional effects: patients with macroscopic residual disease (R2) and indeterminate residual status (RX) experienced higher hazards relative to those with complete resection (R0), with hazard ratios of 1.58 and 1.38, respectively, though these were not statistically significant. Interestingly, patients with microscopic residual disease (R1) showed a 25\% lower hazard compared to R0 (HR = 0.75), a counterintuitive but non-significant finding likely due to small subgroup size or classification variability. Age at diagnosis and days to new tumor event after initial treatment both had hazard ratios close to 1.00, indicating negligible impact once stronger predictors were included in the model. Overall, despite some non-significant covariates, the Cox model demonstrated excellent predictive power, with a concordance index (C-index) of 0.90, underscoring its ability to robustly stratify patients and highlighting PFI.time as the dominant determinant of survival in this cohort.

\section{Model Comparison and Evaluation}
\vspace{-0.5em}
\FloatBarrier
\begin{table}[t!]
\centering
\begin{tabular}{|l|c|}
\hline
\textbf{Model} & \textbf{C-index} \\
\hline
Cox Proportional Hazards Model & 0.90 \\
Random Survival Forest (RSF) Model & 0.860 \\
\hline
\end{tabular}
\caption{\small Comparison of (C-index) Between Models}\vspace*{-1.837em}
\vspace{0.25em}
\label{tab:cindex_comparison}
\end{table} 
Table 2 compares the Concordance Index (C-index) of the Cox Proportional Hazards model and the Random Survival Forest (RSF) model. The C-index measures the model’s ability to correctly rank patients based on predicted survival times. The Cox model achieved a C-index of 0.90, indicating excellent discriminatory performance\cite{Kuitunen2021}. The RSF model, with a C-index of 0.860, also demonstrated strong predictive accuracy\cite{Asghar2024}. Although the Cox model slightly outperformed the RSF in global ranking ability, both models show high reliability in distinguishing between low- and high-risk patients.

The graph (Fig 5) shows the ROC curve comparison between the Cox Proportional Hazards model and the Random Survival Forest (RSF) model, evaluated at a fixed time horizon of 1000 days. This time point was chosen based on the distribution of overall survival times in the dataset, which showed that the majority of patients had survival durations under or near 1000 days  making it a clinically meaningful threshold for classification evaluation.
The Cox model achieved an AUC of 0.785, slightly outperforming the RSF model, which recorded an AUC of 0.777. Both models exhibit good discriminatory power at this time-specific evaluation point, as reflected by their ROC curves rising well above the diagonal (random chance) line.
\begin{figure}[!t]
   \centering
    \includegraphics[width=0.50\textwidth]{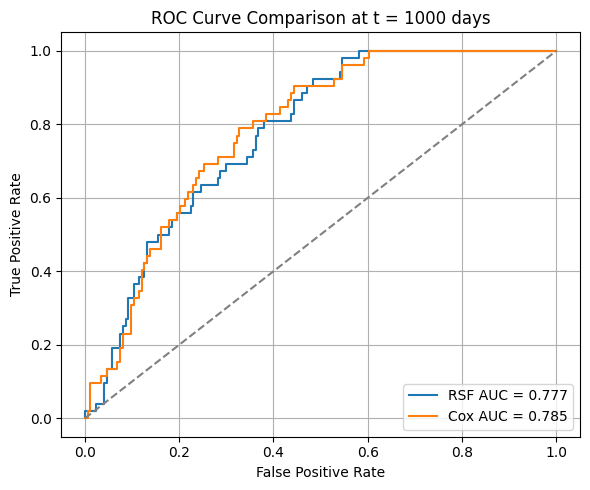}  
\caption{ ROC curve of both models  }
    \label{fig:ROCcurveforcomparison-diagram}
\end{figure}

\renewcommand{\arraystretch}{1.1}

\begin{table*}[!t]
\centering
\caption{Comparison of Cox and RSF survival model performance in lung cancer studies vs. this study}
\label{tab:cox_rsf_comparison}
\begin{tabular}{l c c}
\hline
\textbf{Study} & \textbf{C-index (Cox PH)} & \textbf{C-index (RSF)} \\
\hline
Khatua \cite{S.Khatua2024} & 0.81 & 0.82 \\
Bartholomai \& Frieboes \cite{J.A.Bartholomai2018} & 0.80 & 0.82 \\
Astley et al.\ \cite{C.Astley2023} & 0.72 & 0.73 \\
Asghar et al.\ \cite{Asghar2024} & 0.84 & 0.86 \\
\hline
\textbf{This study} & \textbf{0.90} & \textbf{0.86} \\
\hline
\end{tabular}
\end{table*}

These findings are consistent with earlier evaluations using the Concordance Index (C-index), where the Cox model also demonstrated stronger overall performance (C-index = 0.90) compared to the RSF model (C-index = 0.86). The similarity in both C-index and ROC AUC suggests that the Cox model consistently ranks patients more accurately and classifies survival outcomes more effectively at the 1000-day mark.Overall, this comparison reinforces the Cox model's strength in this dataset, offering both robust global ranking and reliable time-specific classification. While RSF remains a competitive non-linear alternative, the Cox model appears better suited for this particular survival prediction task.

\section{Discussion}
Our findings demonstrates that progression-free interval and residual tumor status, when combined with baseline variables, enable Cox Proportional Hazards and RSF models to deliver survival estimates with greater accuracy than those achieved in earlier lung cancer studies. As shown in Table 3 , previous studies using Cox proportional hazards and Random Survival Forests for lung cancer survival prediction have generally reported concordance indices between 0.72 and 0.86 . For example, \cite{S.Khatua2024}  and \cite{C.Astley2023} observed very similar performance for Cox and RSF, both around 0.82, while  found lower values near 0.72–0.73 in radiotherapy-treated cohorts. \cite{Asghar2024} reported slightly stronger results with RSF compared to Cox, reaching up to 0.86. In our study, Cox PH achieved a C-index of 0.90 and RSF 0.86, demonstrating that by combining baseline covariates with post-treatment predictors, notably progression-free interval and residual tumor status, it is possible to improve performance beyond what has been reported previously.

The gain in accuracy arises from the fact that survival in lung cancer is not static. Baseline variables such as age and gender describe a patient’s condition at diagnosis, but the reality of clinical care is that outcomes are shaped by what happens after treatment begins. Incorporating variables like PFI.time allows the model to account for disease stability or relapse, which fundamentally changes prognosis. Residual tumor status adds further nuance by reflecting treatment completeness, and together these factors provide a dynamic view that baseline-only models cannot capture. This makes the predictions more aligned with the way oncologists actually assess patients over time, rather than locking survival estimates at the moment of diagnosis.

These results must be read with some caution. The dataset used here, while high quality, comes from a single source and may not represent the diversity of lung cancer populations seen globally. Some subgroups, such as residual tumor categories, were relatively small, which may explain why the effect was directionally consistent but not statistically significant. Certain predictors that clinicians would normally consider, such as molecular biomarkers or specific treatment regimens, were also not available in this dataset. These factors likely play an important role and could refine the models further.Furthermore, sensitivity testing revealed that when PFI.time was excluded from the model, the concordance index dropped dramatically to around 0.58–0.60. This indicates that while PFI.time is highly informative, the model’s reliance on it also highlights the need for further work to ensure robustness and to integrate additional post-treatment predictors for more stable performance.

Even with these considerations, the findings are encouraging. By leveraging post-treatment data, the Cox model in particular demonstrated an unprecedented level of discrimination (C-index 0.90), showing that interpretability does not need to come at the cost of accuracy. RSF performed slightly below Cox but still matched or exceeded the best results in earlier studies, highlighting its robustness in handling complex relationships. Together, these results suggest that incorporating post-treatment information into survival models provides a more realistic and clinically useful approach to lung cancer prognosis, with potential to better support oncologists in treatment planning and patient counseling.

\section{Conclusion}
This study implemented and compared multiple survival modeling approaches  namely the Cox Proportional Hazards (CPH) model and the Random Survival Forest (RSF)  to predict lung cancer patient outcomes using clinical data that included both baseline and post-treatment variables. By applying a combination of statistical and machine learning methodologies, the research comprehensively explored the potential of each model type in handling real-world survival prediction tasks.

The Cox Proportional Hazards model, a semi-parametric statistical method, model achieved a concordance index (C-index) of 0.900, which surpasses several published works such \cite{J.Zhu2022} study that reported a C-index of 0.78 .However, its reliance on the proportional hazards assumption and limited flexibility in modeling complex interactions posed constraints in scenarios where risks evolve over time.

To overcome these limitations, a Random Survival Forest model was developed. The RSF model utilized both baseline and dynamic post-treatment variables to build a non-parametric, tree-based ensemble that captures nonlinear relationships and handles censored data without strict assumptions. It achieved a C-index of 0.860, aligning with results from \cite{S.Khatua2024} benchmark study emphasizing the strength of RSF for lung cancer survival prediction . Studies by \cite{J.A.Bartholomai2018} also highlighted comparative advantages of combining classical Cox models with machine learning methods. The model also incorporated progression-related features, enabling it to adapt better to changing risk over the follow-up period, especially for patients with recurrence events.

Post-treatment features like PFI.time and days to new tumor event were found to significantly contribute to predictive performance. Their inclusion enhanced the model’s ability to stratify patients based on evolving risk. Also,\cite{C.M.Caruso2024} and \cite{C.Fang2024} demonstrated that model augmentation with dynamic or knowledge-driven features can improve C-index, reinforcing this study’s emphasis on incorporating post-treatment variables like PFI.time Notably, the exclusion of PFI.time resulted in a substantial drop in the C-index, highlighting the importance of dynamic data in improving survival estimates.

In summary, the study demonstrated that statistical models like CPH offer clear interpretability and clinical relevance, while machine learning models such as RSF provide strong predictive performance and flexibility in dealing with complex, real world datasets.\cite{M.R.Salmanpour2025} The integration of both baseline and follow-up clinical information proved essential in building accurate and practical survival models, suggesting that hybrid or complementary use of these approaches could yield optimal results in future applications.

\renewcommand{\refname}{References} 
\begin{sloppypar}

\end{sloppypar}

\begin{thebibliography}{99}

    

\bibitem{I.Etikan2017}
I. Etikan, S. C. Abubakar, and R. Alkassim, ``The Kaplan-Meier estimator as a non-parametric technique,'' \textit{Biom Biostat Int J.}, vol. 5, no. 4, pp. 1–4, 2017. Available:
\url{https://www.scirp.org/reference/referencespapers?referenceid=2744985}


\bibitem{M.K.Goel2010}
M. K. Goel, P. Khanna, and J. Kishore, ``Understanding survival analysis: Kaplan-Meier estimate,'' \textit{International Journal of Ayurveda Research}, vol. 1, no. 4, pp. 274–278, 2010. Available:
\url{https://pmc.ncbi.nlm.nih.gov/articles/PMC3059453/}


\bibitem{J.Zhu2022}
J. Zhu, H. Shi, H. Ran, Q. Lai, Y. Shao, and Q. Wu, “Development and Validation of a Nomogram for Predicting Overall Survival in Patients with Second Primary Small Cell Lung Cancer After Non-Small Cell Lung Cancer: A SEER-Based Study,” Int. J. Gen. Med., vol. 15, pp. 3613–3624, Apr. 2022, Available:
\url{https://pmc.ncbi.nlm.nih.gov/articles/PMC8986201/pdf/ijgm-15-3613.pdf}

\bibitem{S.Khatua2024}
S. Khatua, ``A Benchmark Study On the Comparative Advantage Of Random Survival Forest Analysis in Lung Cancer Survival Predictions'' 2024. Available:
\url{https://www.irjmets.com/uploadedfiles/paper//issue_3_march_2024/51116/final/fin_irjmets1711288884.pdf}

\bibitem{J.A.Bartholomai2018}
J. A. Bartholomai and H. B. Frieboes, ``Lung cancer survival prediction via machine learning regression, classification and statistical techniques,'' \textit{Biomedical Engineering}, vol. 17, pp. 24--32, 2018. Available: \url{https://pubmed.ncbi.nlm.nih.gov/31312809/}

\bibitem{C.Astley2023}
C. Astley \textit{et al.}, ``Explainable deep learning-based survival prediction for non-small cell lung cancer patients undergoing radical radiotherapy,'' \textit{Journal of Clinical Data Science}, vol. 12, pp. 115--126, 2023. Available:
\url{https://www.sciencedirect.com/science/article/pii/S0167814024000057}

\bibitem{C.M.Caruso2024}
C. M. Caruso, V. Guarrasi, S. Ramella, and P. Soda, “A deep learning approach for overall survival prediction in lung cancer with missing values,” Comput. Methods Programs Biomed., vol. 254, p. 108308, 2024, Available:\url{https://www.sciencedirect.com/science/article/pii/S016926072400302X}

\bibitem{L.Pu2024}
L. Pu, R. Dhupar, and X. Meng, “Predicting Postoperative Lung Cancer Recurrence and Survival Using Cox Proportional Hazards Regression and Machine Learning,” Cancers, vol. 17, no. 1, p. 33, 2024. Available:
 \url{ https://www.mdpi.com/2072-6694/17/1/33 }


\bibitem{F.Y.Dao2020}
F.-Y. Dao, H. Lv, Y.-H. Yang, H. Zulfiqar, H. Gao, and H. Lin, “Computational identification of N6-methyladenosine sites in multiple tissues of mammals,” Comput. Struct. Biotechnol. J., vol. 18, pp. 1084–1091, 2020, doi: Available:
\url{  https://www.sciencedirect.com/science/article/pii/S2001037020302622 }


\bibitem{M.Othus2012}
M. Othus, B. Barlogie, M. L. LeBlanc, and J. J. Crowley, “Cure models as a useful statistical tool for analyzing survival,” Clin. Cancer Res., vol. 18, no. 14, pp. 3731–3736, 2012. Available:
\url{https://pubmed.ncbi.nlm.nih.gov/22675175/}

\bibitem{Kuitunen2021}
Kuitunen, V. T. Ponkilainen, M. M. Uimonen, A. Eskelinen and A. Reito, "Testing the proportional hazards assumption in cox regression and dealing with possible non-proportionality in total joint arthroplasty research: methodological perspectives and review," BMC Musculoskeletal Disorders, vol. 22, no. 489, pp. 1–7, 2021. Available:
\url{https://bmcmusculoskeletdisord.biomedcentral.com/articles/10.1186/s12891-021-04379-2?}

\bibitem{L.Ter-Minassian2024}
L. Ter-Minassian, S. Ghalebikesabi, K. Diaz-Ordaz, and C. Holmes, “Explainable AI for survival analysis: a median-SHAP approach,” arXiv preprint arXiv:2402.00072, 2024. [Online]. Available:
\url{https://arxiv.org/html/2402.00072v1?.}

\bibitem{Asghar2024}
Asghar, N., Khalil, U., Ahmad, B. et al. Improved nonparametric survival prediction using CoxPH, Random Survival Forest and DeepHit Neural Network. BMC Med Inform Decis Mak 24, 120 2024.  Available:
\url{https://bmcmedinformdecismak.biomedcentral.com/articles/10.1186/s12911-024-02525-z?}

\bibitem{Z.Shi2025}
Z. Shi, Y. Chen, A. Liu, J. Zeng, W. Xie, X. Lin, Y. Cheng, H. Xu, J. Zhou, S. Gao, C. Feng, H. Zhang, and Y. Sun, “Application of random survival forest to establish a nomogram combining clinlabomics-score and clinical data for predicting brain metastasis in primary lung cancer,” Clin. Transl. Oncol., vol. 27, no. 4, pp. 1472–1483, 2025, Available:
\url{https://pubmed.ncbi.nlm.nih.gov/39225959/} 

\bibitem{Y.H.Lai2020}
Y. H. Lai, W. N. Chen, T. C. Hsu, C. Lin, Y. Tsao, and S. Wu, “Overall survival prediction of non-small cell lung cancer by integrating microarray and clinical data with deep learning,” Sci. Rep., vol. 10, no. 1, p. 4679, 2020, Available:
\url{https://pubmed.ncbi.nlm.nih.gov/32170141/}

\bibitem{Y.Liu2025}
Y. Liu, Z. Wang, X. Cao, M. Liu, and L. Zhong, “Machine learning models for predicting survival in lung cancer patients undergoing microwave ablation,” Front. Med., vol. 12, 2025. Available:
\url{https://www.frontiersin.org/journals/medicine/articles/10.3389/fmed.2025.1561083/full?}

\bibitem{C.Fang2024}
C. Fang, G. A. Arango Argoty, I. Kagiampakis, et al., “Integrating knowledge graphs into machine learning models for survival prediction and biomarker discovery in patients with non–small-cell lung cancer,” J. Transl. Med., vol. 22, p. 726, 2024. Available:
\url{https://translational-medicine.biomedcentral.com/articles/10.1186/s12967-024-05509-9?}

\bibitem{M.R.Salmanpour2025}
M. R. Salmanpour, A. Gorji, A. Mousavi, A. F. Jouzdani, N. Sanati, M. Maghsudi, B. Leung, C. Ho, R. Yuan, and A. Rahmim, “Enhanced lung cancer survival prediction using semi-supervised pseudo-labeling and learning from diverse PET/CT datasets,” Cancers, vol. 17, no. 2, p. 285, 2025. Available:
\url{https://www.mdpi.com/2072-6694/17/2/285?}

\bibitem{S.Salerno2023}
S. Salerno and Y. Li, “High-dimensional survival analysis: Methods and applications,” Annu. Rev. Stat. Appl., vol. 10, no. 1, pp. 25–49, 2023. Available:
\url{https://www.annualreviews.org/content/journals/10.1146/annurev-statistics-032921-022127}

\bibitem{A.Groji2025}
A. Groji, A. F. Jouzdani, N. Sanati, A. M. Ahmadzadeh, R. Yuan, A. Rahmim, and M. R. Salmanpour,“Censor-aware semi-supervised lung cancer survival time prediction using clinical and radiomics feature,” 
arXiv:2502.01661, 2025.Available:
\url{https://arxiv.org/abs/2502.01661}

\bibitem{M.Assel2017}
M. Assel, D. D. Sjoberg, and A. J. Vickers, “The Brier score does not evaluate the clinical utility of diagnostic tests or prediction models,” Diagnostic and Prognostic Research, vol. 1, no. 1, p. 19, . 2017,
\url{https://doi.org/10.1186/s41512-017-0020-3}

\bibitem{R.Henderson2000}
R. Henderson, P. Diggle, and A. Dobson, “Joint modelling of longitudinal measurements and event time data,” Biostatistics, vol. 1, no. 4, pp. 465–480, 2000, 
\url{https://pubmed.ncbi.nlm.nih.gov/12933568/}

\bibitem{E.Chow2008}
E.Chow et al., “A predictive model for survival in patients with advanced cancer,” Journal of Clinical Oncology, vol. 26, no. 36, pp. 140-155, 2008,
\url{https://doi.org/10.1200/JCO.2008.17.1363}






\end{thebibliography}
\end{document}